\documentclass[twocolumn,showpacs,superscriptaddress,preprintnumbers,prd]{revtex4-1}
\usepackage{graphicx,amsmath,amssymb,dcolumn,bm}
\usepackage{fixmath}
\usepackage{feynmp}
\usepackage{epic,eepic}
\usepackage{slashed}

\begin{document}
\preprint{KEK-TH-2192, J-PARC-TH-0215}
\title{Deuteron polarizations in the proton-deuteron Drell-Yan process \\
for finding the gluon transversity}
\author{S. Kumano}
\email[]{shunzo.kumano@kek.jp}
\affiliation{KEK Theory Center,
             Institute of Particle and Nuclear Studies, \\
             High Energy Accelerator Research Organization (KEK), \\
             Oho 1-1, Tsukuba, Ibaraki, 305-0801, Japan}
\affiliation{J-PARC Branch, KEK Theory Center,
             Institute of Particle and Nuclear Studies, KEK, \\
           and Theory Group, Particle and Nuclear Physics Division, 
           J-PARC Center, \\
           Shirakata 203-1, Tokai, Ibaraki, 319-1106, Japan}
\affiliation{Department of Particle and Nuclear Physics, \\
             Graduate University for Advanced Studies (SOKENDAI), \\
             Oho 1-1, Tsukuba, Ibaraki, 305-0801, Japan} 
\author{Qin-Tao Song}
\email[]{songqintao@zzu.edu.cn}
\affiliation{School of Physics and Microelectronics, Zhengzhou University, \\
             Zhengzhou, Henan 450001, China}
\date{April 28, 2020}

\begin{abstract}
The gluon transversity distribution in the deuteron is defined 
by the matrix element between linearly-polarized deuteron states,
and it could be investigated in proton-deuteron collisions 
in addition to lepton-deuteron scattering.
The linear polarization of photon is often used, whereas it is rarely 
used for the spin-1 deuteron. Therefore, it is desirable to express 
deuteron-reaction cross sections in term of conventional deuteron 
spin polarizations for actual experimental measurements. In this work, 
we investigate how proton-deuteron Drell-Yan cross sections are 
expressed by the conventional polarizations for finding 
the gluon transversity distribution.
In particular, we show that the gluon transversity can be measured
in the proton-deuteron Drell-Yan process by taking the cross-section
difference between the deuteron spin polarizations 
along the two-transverse axes.
Since the gluon transversity does not exist for the spin-1/2 nucleons,
a finite gluon transversity of the deuteron could indicate an ``exotic"
mechanism beyond the simple bound system of the nucleons in the deuteron.
Therefore, the gluon transversity is an interesting and appropriate 
observable to shed light on a new hadronic mechanism in nuclei.
\end{abstract}
\maketitle

\section{Introduction}
\label{intro}

The nucleon spin used to be explained by a combination of
three-quark spins in the nucleon according to the basic quark model.
However, it became clear that the quark contribution 
to the nucleon spin is 20$-$30\%, and the rest of spin should 
come from gluon-spin and partonic orbital-angular-momentum (OAM) 
contributions \cite{nucleon-spin}.
Now, the controversial issue has basically settled down on the nucleon-spin
decomposition into partonic components of spin and OAM contributions
in the color-gauge invariant matter \cite{decomposition}.
Recently, major efforts have been done in lattice quantum chromodynamics (QCD)
\cite{lattice-pdfs} and experimental measurements to find each component, 
and such projects will continue in the next decade.

In particular, three-dimensional structure functions need to be
studied for determining the OAM contributions. Among the three-dimensional
structure functions, the generalized parton distributions (GPDs) 
are used for finding the OAM components by integrals
over the Bjorken scaling variable \cite{gpds-gdas,gpds}.
Furthermore, generalized distribution amplitudes (GDAs), 
which are the $s$-$t$ crossed quantities of the GPDs, 
are also valuable for investigating internal structure 
of hadrons \cite{gdas}. Here, $s$ and $t$ are 
the Mandelstam variables. For probing the transverse structure of hadrons,
there are also transverse-momentum-dependent parton distributions (TMDs) as 
other three-dimensional structure functions \cite{tmds}.

So far, spin structure of the nucleons has been investigated mainly
by longitudinal-polarization observables; however, the transverse spin
measurements started to appear. 
One of important transverse-polarization
quantities is the quark transversity \cite{br-book}. 
Quark transversity distributions are defined by helicity-flip 
amplitudes for quarks, so that they are often called chiral-odd 
distributions. There are some studies toward a global analysis 
on their determination; however, it is still at a preliminary stage 
in the sense that the obtained distributions have large errors
\cite{transversity-pdfs}. 

The gluon transversity distribution is expressed by the amplitude
of the gluon helicity flip ($\Delta s =2$)
\cite{gluon-trans-th,transversity-model,transversity-lattice}.
Therefore, the gluon transversity does not exist for the spin-1/2 nucleon 
due to the helicity-conservation constraint. 
A hadron with spin more than or equal to one is needed 
for the helicity flip of two units.
It means that the scaling violation of the quark transversity
distributions of the nucleon is much different from the one of 
the longitudinally-polarized ones because they are decoupled
from the gluon distribution
\cite{transversity-q2,transversity-q2-gluon}.
It should be noted that the name ``gluon transversity" is misleading 
because it is not related to the transverse polarization 
of the gluon and it is defined by linear polarizations.
The gluon transversity distribution is used 
for the gluon collinear distribution connected to 
the difference between the linear polarizations along 
the transverse axes $x$ and $y$.

The most simple and stable target for studying the gluon transversity
is the deuteron. The deuteron is basically a bound state of 
a proton and a neutron. However, the spin-1/2 nucleons cannot
contribute directly to the gluon transversity distribution, so that
it is an appropriate observable for finding an ``exotic" component
in a nucleus beyond a simple bound system of the nucleons
\cite{transversity-model}.
An exotic component in the deuteron could be also investigated by 
tensor-polarized structure functions, for example $b_1$
\cite{b1}, by considering that there are significant differences
between conventional deuteron calculations \cite{b1-convolution}
and the HERMES measurement \cite{b1-hermes}. 
Such studies will be done by the approved 
Thomas Jefferson National Accelerator Facility (JLab) experiment
\cite{Jlab-b1} and possibly by a Fermilab experiment
\cite{Fermilab-dy,ks-2016}
in the measurements of $b_1$ and the tensor-polarized parton distribution
functions (PDFs), respectively. 
However, a unique point of the gluon transversity is that
the direct nucleon contribution does not exist (namely $\Delta_T g =0$), 
whereas the direct contribution is finite in $b_1$ and 
the tensor-polarized PDFs, for example, due to the D-wave component 
in the proton-neutron bound system \cite{b1-convolution}.
Any finite gluon transversity distribution ($\Delta_T g \ne 0$)
could indicate an interesting new mechanism beyond the bound state 
of proton and neutron.

At this stage, there is no experimental measurement on the gluon transversity
distribution. The only possibility is the letter of intent to measure 
the gluon transversity at JLab by using the polarized spin-1 
deuteron \cite{jlab-gluon-trans}. 
For example, by finding the azimuthal-angle 
dependence of the deuteron transverse polarization, 
the gluon transversity distribution will be measured. 
So far, it is the only experimental project and such an effort 
will be continued at Electron-Ion Collider (EIC) \cite{Jlab-b1}.
On the other hand, there are a number of hadron accelerators in the world, 
and it is excellent if similar measurements will be done
at the hadron facilities. 
There are available hadron facilities at Fermilab \cite{Fermilab-dy}, 
Japan Proton Accelerator Research Complex (J-PARC) \cite{j-parc}, 
Gesellschaft f\"ur Schwerionenforschung -Facility 
for Antiproton and Ion Research (GSI-FAIR) \cite{gsi-fair}, and 
Nuclotron-based Ion Collider fAcility (NICA) \cite{NICA-SPD}.
In addition, if the fixed-deuteron target becomes possible
at Relativistic Heavy Ion Collider (RHIC) \cite{RHIC-fixed}, 
Large Hadron Collider (LHC), or EIC, there is another possibility. 
In general, lepton- and hadron-accelerator measurements are complementary
and both results are essential in establishing hadron structure
in a wide kinematical region.

For such a purpose, we proposed to use the proton-deuteron
Drell-Yan process with the polarized deuteron for finding
the gluon transversity distribution in the deuteron
\cite{ks-gluon-trans-2019}. The cross-section formulae
and numerical values were shown for the cross sections
by taking linear polarizations of the deuteron.
The linear polarization is often used in photon physics.
It is also known that the linear-polarization states of the photon 
are expressed by the circular-polarization states 
with the helicities $+1$ and $-1$ \cite{photon-polarization,leader-book}.
Similarly, the linear polarizations of the deuteron could
be expressed by its helicity or transversely-polarized states.
The proton-deuteron Drell-Yan cross sections are shown by
the linear polarizations in Ref.\,\cite{ks-gluon-trans-2019}.
Such a formalism is theoretically appropriate. However, it is 
more practical to express the cross sections
in terms of conventional spin polarizations because 
the linear polarizations are rarely used in deuteron experiments.

In fact, the electron-scattering cross section is expressed
by the azimuthal angle of the transversely-polarized deuteron
for probing the gluon transversity in the JLab experiment 
\cite{jlab-gluon-trans}. 
In the same way, we expect that the proton-deuteron reaction 
cross sections could be expressed by the transverse polarizations 
of the deuteron. The purpose of this work is to express 
the proton-deuteron Drell-Yan cross sections in terms of 
the conventional polarizations of the deuteron 
for future experimental measurements.

In this article, we explain deuteron polarizations,
their rotations around the transverse axes, and
collinear parton correlation functions of the deuteron
in Sec.\,\ref{d-polarization}.
Next, we show how the proton-deuteron Drell-Yan cross sections 
are expressed by the conventional deuteron polarizations
for probing the gluon transversity
in Sec.\,\ref{pd-dy-cross-section}.
The results are summarized in Sec.\,\ref{summary}.

\section{Deuteron polarizations}
\label{d-polarization}

The vector and tensor polarizations ${\mathbold S}$ 
and ${\mathbold T}$ of the deuteron are expressed by possible
polarization parameters and polarization vector $\vec E$
in Sec.\,\ref{d-polarization-correlation}.
Then, the collinear parton correlation functions of the deuteron
are written by the polarization parameters and the deuteron PDFs.
Next, rotations of the polarization and spin vectors around
the transverse axes are discussed in Sec.\,\ref{rotations-helicity-1}
for relating them with the gluon transversity distribution
in the proton-deuteron Drell-Yan cross sections.

\subsection{Deuteron polarizations \\ and parton correlation functions}
\label{d-polarization-correlation}

The spin vector ${\mathbold S}$ and tensor ${\mathbold T}$
of the deuteron are parametrized in the deuteron rest frame as
\cite{ks-gluon-trans-2019,bacchetta-2000-PRD,vonDaal-2016,Boer-2016}
\begin{align}
\! \! \! \! 
{\mathbold S} & = (S_{T}^x,\, S_{T}^y,\, S_L) ,
\nonumber \\
\! \! \! \! 
{\mathbold T}  & = \frac{1}{2} 
\left(
    \begin{array}{ccc}
     - \frac{2}{3} S_{LL} + S_{TT}^{xx}    & S_{TT}^{xy}  & S_{LT}^x  \\[+0.20cm]
     S_{TT}^{xy}  & - \frac{2}{3} S_{LL} - S_{TT}^{xx}    & S_{LT}^y  \\[+0.20cm]
     S_{LT}^x     &  S_{LT}^y              & \frac{4}{3} S_{LL}
    \end{array}
\right) ,
\label{eqn:st-1}
\end{align}
where $S_{T}^x$, $S_{T}^y$, $S_L$, 
$S_{LL}$, $S_{TT}^{xx}$, $S_{TT}^{xy}$, $S_{LT}^x$, and $S_{LT}^y$
are the parameters to indicate the deuteron's vector 
and tensor polarizations.
The deuteron polarization vector $\vec E$ is defined as
\begin{align}
& \vec E_\pm =\frac{1}{\sqrt{2}} \left ( \, \mp 1,\, -i,\, 0 \, \right ) ,  
\ 
\vec E_0    = \left ( \, 0,\, 0,\, 1 \, \right ) ,
\nonumber \\
& \vec E_x  =\frac{1}{\sqrt{2}} \left ( \vec E_- - \vec E_+ \right )
    = \left ( \, 1,\, 0,\, 0 \, \right ) ,
\nonumber \\
& \vec E_y  =  \frac{i}{\sqrt{2}} \left ( \vec E_- + \vec E_+ \right )
    = \left ( \, 0,\, 1,\, 0 \, \right ) ,
\label{eqn:dct}
\end{align}
where $\vec E_+$, $\vec E_0$, and $\vec E_-$
indicate the spin states with the $z$ component of spin
$s_z =+1$, $0$, and $-1$.
The polarizations $\vec E_x$ and $\vec E_y$ are called linear polarizations,
and they are necessary for investigating the gluon transversity
distribution in the deuteron \cite{ks-gluon-trans-2019}.
The gluon transversity distribution is defined by
the matrix element between the linearly-polarized deuteron states.
In terms of the polarization vector $\vec E$ of the deuteron,
they are written as \cite{ks-gluon-trans-2019,leader-book}
\begin{align}
{\mathbold S}
= \text{Im} \, (\, \vec E^{\, *} \times \vec E \,),
\ \ \ 
T_{ij}  = \frac{1}{3} \delta_{ij} 
       - \text{Re} \, (\, E_i^{\, *} E_j \,) .
\label{eqn:spin-1-vector-tensor-2}
\end{align}

For example, if the polarization vector is $\vec E_+ = (-1,\, -i,\, 0)/\sqrt{2}$,
we obtain the vector and tensor polarizations from 
Eq.\,(\ref{eqn:spin-1-vector-tensor-2}) as
\begin{align}
{\mathbold S}  = (0,\, 0,\, 1) , \ \ \ 
{\mathbold T}  = 
\left(
    \begin{array}{ccc}
     - \frac{1}{6}     & 0  & 0 \\[+0.20cm]
   0  & - \frac{1}{6}     & 0 \\[+0.20cm]
     0     &  0           & \frac{1}{3}
    \end{array}
\right) .
\label{eqn:st-2}
\end{align}
In comparison with Eq.\,(\ref{eqn:st-1}), these relations indicate 
the polarization parameters as
\begin{align}
S_L & =1, \ \  S_{LL}=1/2, 
\nonumber \\
S_{T}^x & = S_{T}^y = S_{TT}^{xx} 
= S_{TT}^{xy} = S_{LT}^x =S_{LT}^y =0 .
\label{eqn:spin-parameters+1}
\end{align}
The polarization $\vec E_+$ means that the deuteron polarization along $z$
($S_L=1$), and it also contains a part of the tensor-polarization $S_{LL}$
according to Eq.\,(\ref{eqn:spin-parameters+1}).

Deuteron-reaction cross sections are expressed generally 
by correlation functions. The twist-2 collinear quark-correlation function 
for the spin-1 deuteron, which is denoted as $B$,
is given \cite{ks-gluon-trans-2019,bacchetta-2000-PRD,vonDaal-2016}
in the laboratory coordinates as
\begin{align}
\Phi_{q/B} & (x_b) 
 = \frac{1}{2} \bigg [  \, 
\slashed{\bar n}  \, f_{1,q/B} (x_b) 
+ \gamma_5 \, \slashed{\bar n} \, S_{B,L} \, g_{1,q/B} (x_b)
\nonumber \\
& 
+ \slashed{\bar n} \, \gamma_5 \, \slashed{S}_{B,T} \, h_{1,q/B} (x_b)
  + \slashed{\bar n} \, S_{B,LL} \, f_{1 LL, q/B} (x_b)
\nonumber \\
& 
+ \sigma_{\mu\nu} \, {\bar n} ^\nu \, S_{B,LT}^\mu \, h_{1LT,q/B} (x_b) \,
\bigg ] ,
\label{eqn:correlation-integrated-q-deuteron} 
\end{align}
where $f_{1,q/B} (x_b)$ is the unpolarized quark-distribution function,
$g_{1,q/B} (x_b)$ is the longitudinally-polarized one,
$h_{1,q/B} (x_b)$ ($= \Delta_T q_B(x_b)$)
is the transversity, 
and $f_{1 LL, q/B} (x_b)$  
and $h_{1LT,q/B} (x_b)$ 
are tensor-polarized ones \cite{bacchetta-2000-PRD}.
The variable $x_b$ is the momentum fraction carried by a parton
in the deuteron. The lightlike vectors $\bar n$ and $n$ are defined by
\begin{align}
\bar n^\mu  = \frac{1}{\sqrt{2}}  (\, 1,\, 0,\, 0,\, +1\, ) , \ \ 
     n^\mu  = \frac{1}{\sqrt{2}}  (\, 1,\, 0,\, 0,\, -1\, ) ,
\label{eqn:n-nbar-2}
\end{align}
and the antisymmetric tensor $\sigma_{\mu\nu}$ is given by
$\sigma_{\mu\nu} = \frac{i}{2} (\gamma_\mu \gamma_\nu -\gamma_\nu \gamma_\mu )$.

The twist-2 gluon correlation function is similarly given 
for the deuteron as \cite{ks-gluon-trans-2019,vonDaal-2016,Boer-2016}
\begin{align}
& \! \! \! \! 
\Phi_{g/B}^{\, \alpha\beta} (x_b) 
= \frac{1}{2} \bigg [ - g_T^{\, \alpha\beta} f_{1,g/B} (x_b) 
+ i \, \epsilon_T^{\, \alpha\beta} S_{B,L} \, g_{1,g/B} (x_b)
\nonumber \\
& \! \! 
- g_T^{\, \alpha\beta} S_{B,LL} f_{1LL,g/B} (x_b) 
+ S_{B,TT}^{\, \alpha\beta} \, h_{1TT,g/B} (x_b)
\, \bigg ] .
\label{eqn:correlation-integrated-g} 
\end{align}
Here, $f_{1,g/B}$ is the unpolarized gluon distribution function,
$g_{1,g/B}$ is the longitudinally-polarized one,
$f_{1 LL, g/B}$ and $h_{1TT,g/B}$ 
($\equiv - \Delta_T g_B$ in this paper)
are tensor- and linearly-polarized ones. 
In Eq.\,(\ref{eqn:correlation-integrated-g}),
$g_T^{\alpha\beta}$ is defined by 
$g_T^{\alpha\beta} = g^{\alpha\beta} - \bar n^{\{ \alpha} n^{\beta \}}$
($g_T^{11}=g_T^{22}=-1$, $\text{others}=0$)
with the symmetrization 
$a^{\{ \alpha} b^{\beta \}} \equiv a^\alpha b^\beta + a^\beta b^\alpha$
for the superscript indices, 
and $\epsilon_T^{\alpha\beta}$ is given by
$\epsilon_T^{\alpha\beta} \equiv \epsilon^{\alpha\beta - +}$
($\epsilon_T^{12}=-\epsilon_T^{21}=1$, $\text{others}=0$).

\begin{figure}[t]
 \vspace{-0.00cm}
\begin{center}
   \includegraphics[width=8.0cm]{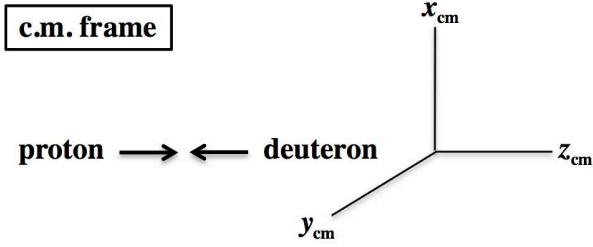}
\end{center}
\vspace{-0.7cm}
\caption{Proton-deuteron reactions in the center-of-momentum frame,
where the deuteron moves in the $-z$ direction.}
\label{fig:cm-pd}
\vspace{-0.00cm}
\end{figure}

The spin vector and tensor in other frames, for example
the center-of-momentum (c.m.) frame, are expressed also by
the polarization parameters, $S_{T}^x$, $S_{T}^y$, $S_L$, 
$S_{LL}$, $S_{TT}^{xx}$, $S_{TT}^{xy}$, $S_{LT}^x$, and $S_{LT}^y$,
which are defined in the rest frame of the deuteron, namely
the laboratory frame in the proton-deuteron reactions at Fermilab
with the proton beam and the fixed-target deuteron.
Let us consider the relation between the spin vectors 
in the laboratory and c.m. frames.
The spin four-vector and momentum are given in the laboratory frame as
\begin{align}
s^ {\,\mu}_{\text{lab}} & = (0,\, S_{T}^x,\, S_{T}^y,\, S_L) 
= (0,\, \vec s_{\text{lab}} \,) 
, 
\nonumber \\
p^ {\,\mu}_{\text{lab}} & = (M,\, 0,\, 0,\, 0) ,
\label{eqn:spin-05a}
\end{align}
where $M$ is the hadron mass.
The spin and momentum vectors satisfy $s^2=-1$, $p^2=M^2$, 
and $s \cdot p=0$ in any frame.
The spin vector and momentum in the c.m. frame 
are given by the Lorentz transformations as
\begin{align}
s^ {\,\mu} & =  \Lambda^{\mu}_{\ \nu} \,  s^{\,\nu}_{\text{lab}}
= (\gamma \, \beta \, S_L,\, S_{T}^x,\, S_{T}^y,\, \gamma \, S_L) ,
\nonumber \\
p^ {\,\mu} & =  \Lambda^{\mu}_{\ \nu} \, p^{\,\nu}_{\text{lab}}
 = (\gamma \, M,\, 0,\, 0,\, \gamma \,\beta \, M) ,
\nonumber \\
& \ 
\Lambda^{\mu}_{\ \nu}  =
\left(
    \begin{array}{cccc}
     \gamma   &0 &0 &\gamma \, \beta \\[+0.05cm]
     0 & 1&0 & 0  \\[+0.05cm]
     0 &0   &1  & 0  \\[+0.05cm]     
   \gamma \, \beta &  0     &0         & \gamma
    \end{array}
\right) , \ \ \gamma = \frac{1}{\sqrt{1-\beta^2}} ,
\label{eqn:spin-05ab}
\end{align}
where the c.m. frame moves with the velocity $-\beta$ 
in the $z$ direction with respect to the laboratory frame.
The relation between the spin vectors of both frames is given by
\cite{br-book,ks-gluon-trans-2019}
\begin{align}
s^\mu = \left ( \,   
        \frac{\vec p \cdot \vec s _{\text{lab}}}{M},\, 
         \vec s _{\text{lab}} + \frac{\vec p \cdot \vec s _{\text{lab}}}
                       {M \, (M+ p^{0})} \, \vec p
        \, \right ) .
\label{eqn:nucleon-spin}
\end{align}
In the relativistic limit $\beta \to 1$, this relation is written as
\begin{align}
s^ {\,\mu} = S_L \frac{p^{\,\mu}}{M} + s_\perp^{\,\mu}.
\label{eqn:spin-ttab}
\end{align}
Here, the factor $S_L$ is the longitudinal-polarization parameter
and the transverse spin $s_\perp^{\,\mu}$ is expressed by the two 
transverse-polarization parameters $S_T^x$ and $S_T^y$ as
\begin{align}
s_\perp^{\,\mu} = (0,\, S_{T}^x,\, S_{T}^y,\, 0) .
\label{eqn:st-mu}
\end{align}
Since the Lorentz boost is in the $-z$ direction for the deuteron
from the laboratory frame to the c.m. frame as shown
in Fig.\,\ref{fig:cm-pd}, one needs to be careful about the spin parameters 
and the correlation functions.
First, the lightlike vector $\bar n$ in 
Eq.\,(\ref{eqn:correlation-integrated-q-deuteron})
should be replaced by $n$ because of 
$\vec p_B \parallel -z$.
Second, noting the covariant forms of $S^\mu$ and $T^{\mu\nu}$ 
in Eq.\,(35) of Ref.\,\cite{ks-gluon-trans-2019}
and the Lorentz boost along the $-z$ direction, 
we find that two parameters change their signs
($S_L \to - S_L$, $S_{LT} \to - S_{LT}$) and the others stay the same.
However, only the terms $S_{B,LL}$ and $S_{B,TT}^{\, xx}$ contribute
to the Drell-Yan cross sections in Sec.\,\ref{pd-dy-cross-section},
the changes do not affect the following discussions.

\subsection{Rotations of deuteron polarizations\\
around transverse axes}
\label{rotations-helicity-1}

\begin{figure}[t]
 \vspace{-0.00cm}
\begin{center}
   \includegraphics[width=8.0cm]{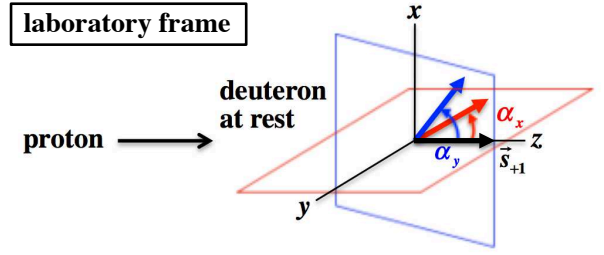}
\end{center}
\vspace{-0.7cm}
\caption{Proton-deuteron reactions in the laboratory frame,
where the polarized deuteron is at rest with spin $\vec s_{+1}$.
The spin vector $\vec s_{+1}$ is rotated with the angle $\alpha_y$ ($\alpha_x$) 
around the axis $y$ ($x$).}
\label{fig:lab-spin}
\vspace{-0.00cm}
\end{figure}

The polarization vectors $\vec E_+$, $\vec E_0$, and $\vec E_-$
of Eq.\,(\ref{eqn:dct}) indicate the spin states $s_z =+1$, $0$, and $-1$ 
by taking the spin quantization axis in the $z$ direction.
The $s_z =0$ state is given by the polarization vector 
$\vec E_0=( \, 0,\, 0,\, 1 \, )$.
In the same way, 
the linear polarizations 
$\vec E_x=( \, 1,\, 0,\, 0 \, )$ and
$\vec E_y=( \, 1,\, 0,\, 0 \, )$ correspond to 
the $s_x =0$ and $s_y =0$ states by taking
the $x$ and $y$ axes, respectively, as the quantization axes,
namely by rotating the $s_z =0$ state around the transverse axes
with the angle $\pi/2$.
If experimentalists can prepare such deuteron-target states, 
the proton-deuteron Drell-Yan cross-section difference
$d\sigma (E_x)-d\sigma (E_y)$ can be measured as suggested
theoretically in Ref.\,\cite{ks-gluon-trans-2019}.
On the other hand, we explain in the following
about the possibility of using the $\vec E_+$ 
($s_z=1$) polarization by rotating the polarization 
around the transverse axes as shown in Fig.\,\ref{fig:lab-spin}.

Since a finite polarization $S_{B,TT}^{\, xx}$ is needed
for finding the gluon transversity $\Delta_T g (x)$ 
in Eq.\.(\ref{eqn:correlation-integrated-g}),
the deuteron state ($\vec E_+$) does not contribute 
for extracting $\Delta_T g (x)$ according to Eq.\,(\ref{eqn:spin-parameters+1}).
Therefore, let us consider a rotation of the polarization vector
$\vec E_{+}$ around the $y$ axis with the angle $\alpha_y$
in the laboratory frame with the deuteron at rest.
Then, the polarization vector $\vec E_{+}$ becomes
\begin{align}
\vec E_{\alpha_y} 
=\frac{1}{\sqrt{2}} \left (-\cos{\alpha_y},\, -i,\,\sin{\alpha_y} \right).
\label{eqn:pola-a} 
\end{align}
Using this polarization vector, we obtain
the spin vector ${\mathbold S}$ and tensor ${\mathbold T}$ as
\begin{align}
\! \! \! \! 
{\mathbold S}_{\alpha_y} & 
= ( \sin{\alpha_y},\, 0,\, \cos{\alpha_y}) ,
\nonumber \\
\! \! \! \! 
{\mathbold T}_{\alpha_y}  & =
\left(
    \begin{array}{ccc}
     \frac{1}{12} \left \{ 1-3\cos (2\alpha_y) \right\}
       &0 & \frac{1}{4}\sin (2\alpha_y) \\[+0.20cm]
     0 & - \frac{1}{6}    & 0  \\[+0.20cm]
    \frac{1}{4}\sin (2\alpha_y)  &  0        
        & \frac{1}{12} \left\{ 1+3\cos(2\alpha_y) \right\}
    \end{array}
\right) .
\label{eqn:st-3}
\end{align}
These vector and tensor polarizations correspond to
the parameter choice:
\begin{alignat}{3}
& S_T^x= \sin{\alpha_y},& \ 
& S_T^y=0,& \ 
& S_L= \cos{\alpha_y},
\nonumber \\ 
& S_{LL}= \frac{1}{8} \left\{ 1+3\cos (2\alpha_y) \right\} ,& \ 
& S_{TT}^{xx}= \frac{1}{2} \sin^2{\alpha_y},&  \ 
& S_{TT}^{xy}=0,
\nonumber \\ 
&  S_{LT}^x= \frac{1}{2} \sin (2\alpha_y),& \ 
&  S_{LT}^y= 0 .&
&   
\label{eqn:vuas1} 
\end{alignat}
If $\alpha_y = \pi/2$ is taken, the polarization vector becomes
\begin{align}
\vec E_{\alpha_y = \pi/2} & = \frac{1}{\sqrt{2}}  \left ( 0,\, -i,\, 1 \right)
               = i \frac{1}{\sqrt{2}}  \left ( 0,\, -1,\, -i \right)
\nonumber \\
       &   \equiv i \vec E_{+/\hat x} ,
\label{eqn:pola-b} 
\end{align}
where $\vec E_{+/\hat x}$ is the polarization vector
with the spin component $+1$ defined by taking the spin-quantization axis 
as $x$.

Next, if the rotation with the angle $\alpha_x$ is applied 
around the $x$ axis, the polarization vector $\vec E_{+}$ becomes 
\begin{align}
\vec E_{\alpha_x} 
& =\frac{i}{\sqrt{2}} \left (i,\, -\cos{\alpha_x},\, -\sin{\alpha_x} \right) .
\label{eqn:pola-c} 
\end{align}
In this case, the polarization parameters are given by
\begin{alignat}{3}
\! \! 
& S_T^x= 0,& \,  
& S_T^y= - \sin{\alpha_x},& \, 
& S_L=  \cos{\alpha_x},
\nonumber \\ 
\! \! 
& S_{LL}= \frac{1}{8} \left\{ 1+3\cos (2\alpha_x) \right\} ,& \, 
& S_{TT}^{xx}= - \frac{1}{2} \sin^2{\alpha_x},&  \, 
& S_{TT}^{xy}=0,
\nonumber \\ 
\! \! 
&  S_{LT}^x= - \frac{1}{2} \sin (2\alpha_x) ,& \, 
&  S_{LT}^y= 0 .&
&   
\label{eqn:vuas2} 
\end{alignat}
At $\alpha_x = - \pi/2$, it corresponds to the spin $+1$ state 
by taking the spin-quantization axis as $y$:
\begin{align}
\vec E_{\alpha_x = -\pi/2} 
& = - i \frac{1}{\sqrt{2}}  \left ( -i,\, 0,\, -1 \right)
   \equiv - i \vec E_{+/\hat y} .
\label{eqn:pola-d} 
\end{align}
In this way, rotating the deuteron spin around the transverse axes,
we obtain a finite linear-polarization parameter $S_{TT}^{xx}$.
Then, the gluon transversity can contribute to 
the gluon correlation function in Eq.\,(\ref{eqn:correlation-integrated-g})
and then to the cross section. However, other spin parameters become
finite according to Eqs.\,(\ref{eqn:vuas1}) and (\ref{eqn:vuas2}),
the deuteron-polarization combination should be appropriately taken
for extracting the gluon transversity distribution as the leading term.
Such a combination will be shown later 
in Eq.\,(\ref{eqn:cross-pd-rotate-plus}).

\section{Polarized proton-deuteron Drell-Yan cross sections \\
for extracting gluon transversity}
\label{pd-dy-cross-section}

In the previous publication \cite{ks-gluon-trans-2019}, the proton-deuteron 
Drell-Yan cross sections were shown for the deuteron linear polarizations 
as $d\sigma (E_x \pm E_y)$.
However, since the linear polarizations are rarely used for the deuteron, 
we need to express them by the usual deuteron polarizations for actual
experimental measurements.
Let us consider the proton-deuteron Drell-Yan process with 
the unpolarized proton and the deuteron polarizations 
$\vec E_{\alpha_y}$ and $\vec E_{\alpha_x}$ or 
the deuteron spin vectors 
${\mathbold S}_{\alpha_y}$, ${\mathbold S}_{\alpha_x}$ 
and tensors
${\mathbold T}_{\alpha_y}$, ${\mathbold T}_{\alpha_x}$.

In the unpolarized proton, there exists one type of twist-2
correlation-function terms, which correspond to the first terms of 
Eq.\,(\ref{eqn:correlation-integrated-q-deuteron}) 
and Eq.\,(\ref{eqn:correlation-integrated-g})
of the deuteron case:
\begin{align}
\Phi_{q/A} (x_a) 
& = \frac{1}{2} \, \slashed{\bar n}  \, f_{1,q/A} (x_a) ,
\nonumber \\
\Phi_{g/A}^{\, \alpha\beta} (x_a) 
& = - \frac{1}{2} g_T^{\, \alpha\beta} \, f_{1,g/A} (x_a) ,
\label{eqn:correlation-integrated-qg-unpol-p} 
\end{align}
where the variable $x_a$ is the momentum fraction carried 
by a parton in the proton.
The cross section of the proton-deuteron Drell-Yan process 
from the subprocess of 
$q\, (\bar q) +g \to \gamma^\ast + q \, (\bar q)$ typically
contains the trace terms
\cite{ks-gluon-trans-2019}
\begin{align}
& \! \! \! \! 
\text{Tr} \left [ (4 \,\gamma \text{'s})
\left \{ \Phi_{q/A} (x_a) + \Phi_{\bar q/A} (x_a) \right \}
\cdots (3 \,\gamma \text{'s}) \Phi_{g/B}^{\, \alpha\beta} (x_b) \right ] ,
\nonumber \\
& \! \! \! \! 
\text{Tr} \left [ (4 \,\gamma \text{'s}) \Phi_{g/A}^{\, \alpha\beta} (x_a) 
\cdots 
(3 \,\gamma \text{'s})
\left \{ \Phi_{q/B} (x_b) + \Phi_{\bar q/B} (x_b) \right \}
\right ] .
\label{eqn:typical-trace}
\end{align}
In the same way, the subprocess 
$q + \bar q \to \gamma^* + g $ typically have the trace
\begin{align}
& \! \! \! \!   
\text{Tr} \left [ (3 \,\gamma \text{'s}) \, \Phi_{q(\bar q)/A} (x_a) \cdots 
(3 \,\gamma \text{'s}) \, \Phi_{\bar q (q)/B} (x_b) \right ] .
\label{eqn:typical-trace-2}
\end{align}
In Eq.\,(\ref{eqn:correlation-integrated-qg-unpol-p}),
because there is one $\gamma$ in $\Phi_{q (\bar q)/A} (x_a)$ and 
no $\gamma$ in $\Phi_{g/A}^{\alpha\beta} (x_a)$, possible terms 
in the deuteron correlation functions should contain
no $\gamma$ in $\Phi_{g/B}^{\, \alpha\beta} (x_b)$
and an odd number of $\gamma$ in $\Phi_{q (\bar q)/B} (x_b)$
so that the trace terms become finite.

Therefore, the relevant correlation-function terms, which 
contribute to the Drell-Yan cross section with the unpolarized proton,
become 
\begin{align}
\! \! 
\Phi_{q/B} (x_b) 
& = \frac{1}{2} \slashed{n} \, \bigg [  f_{1,q/B} (x_b) 
  + \, S_{B,LL} \, f_{1 LL, q/B} (x_b) \bigg ] ,
\nonumber \\
\! \!
\Phi_{g/B}^{\, \alpha\beta} (x_b) 
& = \frac{1}{2} \bigg [  - g_T^{\, \alpha\beta} \, 
\left\{ f_{1,g/B} (x_b) 
+ S_{B,LL} \, f_{1LL,g/B} (x_b) \right\}
\nonumber \\
&  \ \ \ \ \ \ \ \ 
- S_{B,TT}^{\, \alpha\beta} \, \Delta_T g_B (x_b)
\, \bigg ] ,
\label{eqn:correlation-qg-deuteron-mod} 
\end{align}
from Eqs.\,(\ref{eqn:correlation-integrated-q-deuteron})
and (\ref{eqn:correlation-integrated-g}).
Here, the lightlike vector $\bar n$ is replaced by $n$
as mentioned in the end of Sec.\,\ref{d-polarization-correlation}.
The $\alpha_y$ and $\alpha_x$ rotations around $y$ and $x$ 
in Fig.\,\ref{fig:lab-spin} indicate the parameter values as given 
in Eqs.\,(\ref{eqn:vuas1}) and (\ref{eqn:vuas2}):
\begin{alignat}{3}
& \! \! \! \! 
\alpha_y \! : S_{B,LL} \! =  \! \frac{1}{8} 
\left\{ 1 \! + \! 3\cos (2\alpha_y) \right\} \! , & \, \, 
& S_{B,TT}^{xx} \! = \! \frac{1}{2} \sin^2{\alpha_y} , &
\nonumber \\
& \! \! \! \! 
\alpha_x \! :  S_{B,LL} \! = \! \frac{1}{8} 
\left\{ 1 \! + \! 3\cos (2\alpha_x) \right\} \! , & \, \, 
& S_{B,TT}^{xx} \! = \!  - \frac{1}{2} \sin^2{\alpha_x} , &
\label{eqn:polarizations-xy}
\end{alignat}
where the $\gamma_5 \, \slashed{n}$ terms in the quark correlation
function and the antisymmetric term $\epsilon_T^{\, \alpha\beta}$
in the gluon correlation function are also removed because
their contributions vanish in the traces.
The gluon transversity is denoted as
$\Delta_T g_B = - h_{1TT,g/B}$ in Ref.\,\cite{ks-gluon-trans-2019}.

In the proton-deuteron Drell-Yan cross section, there are two types 
of cross sections as obvious from 
Eqs.\,(\ref{eqn:correlation-integrated-qg-unpol-p}), 
and (\ref{eqn:correlation-qg-deuteron-mod}):
\begin{align}
\frac{ d \sigma_{pd \to \mu^+ \mu^- X} }{d\tau \, dq_T^2 \, d\phi \, dy}
= \frac{ d \sigma_{0} (S_{B,LL})}{d\tau \, dq_T^2 \, d\phi \, dy}
+ \frac{ d \sigma_{\Delta_T g} (S_{B,TT}^{xx})}{d\tau \, dq_T^2 \, d\phi \, dy} .
\label{eqn:cross-pd}
\end{align}
Here, the variable $\tau$ is defined by the dimuon-mass squared $M_{\mu\mu}^2$
and the center-of-mass-energy squared $s$ as
\begin{align}
\! \! \!
\tau = \frac{Q^2}{s} , \ 
Q^2  = (k_1 + k_2)^2 = M_{\mu\mu}^2 , \ 
s = (p_A + p_B)^2 ,
\label{eqn:tau}
\end{align}
where $k_1$ and $k_2$ are $\mu^-$ and $\mu^+$ momenta,
and $p_A$ and $p_B$ are proton and deuteron momenta
in the proton-deuteron Drell-Yan process,
$p (p_A) + d (p_B) \to \mu^- (k_1)+\mu^+ (k_2) +X$.
The virtual photon momentum is denoted as $q$ ($=k_1+k_2$),
and it is given as
\cite{ks-gluon-trans-2019}
\begin{align}
q  =  ( \, E,\, q_T \cos\phi,\,  q_T \sin\phi,\,  q_L  ) ,
\label{eqn:photon-q}
\end{align}
by the transverse and longitudinal momenta 
$q_T$ ($= |\vec q\, | \sin\theta$) and $q_L$ ($= |\vec q\, | \cos\theta$) 
with the polar and azimuthal angles 
$\theta$ and $\phi$ in the c.m. frame.
The dimuon rapidity $y$ is defined by 
\begin{align}
y = \frac{1}{2} \ln \frac{E+q_L}{E-q_L} 
  = - \ln \left [ \tan (\theta/2)\right ] ,
\label{eqn:rapidity}
\end{align}
in the c.m. frame.
In Eq.\,({\ref{eqn:cross-pd}), the first term 
$d \sigma_{0}(S_{B,LL})/(d\tau \, dq_T^2 \, d\phi \, dy)$ 
is from the subprocess of the unpolarized PDFs in the proton 
with the unpolarized and tensor-polarized PDFs in the deuteron, 
and the second one 
$d \sigma_{\Delta_T g} (S_{B,TT}^{xx})/(d\tau \, dq_T^2 \, d\phi \, dy)$
is from the gluon transversity distribution in the deuteron.

The first term of Eq.\,({\ref{eqn:cross-pd})
is no more than the cross section of Eqs.\,(101) and (102)
in Ref.\,\cite{ks-gluon-trans-2019}; 
however, the unpolarized distributions
should be replaced by unpolarized plus tensor-polarized one
in the deuteron, $ f_{1,X/B}(x_b)+S_{B,LL} \, f_{1LL,X/B} (x_b)$
with $X=q$, $\bar q$, or $g$, 
where $f_{1LL,X/B} (x_b)$ is a tensor-polarized parton distribution,
according to Eq.\,(\ref{eqn:correlation-qg-deuteron-mod}).
The tensor-polarized PDF is related to the one $\delta_T f$
used in Ref.\,\cite{ks-2016} as 
$f_{1LL,X/B} (x_b) = -(2/3) \delta_T f_{X/B} (x_b)$ \cite{bacchetta-2000-PRD}.
However, the factor of 1/2 needs to be multiplied in Eqs.(101) and (102),
where the combination $E_x+E_y$ was taken, 
and this factor appears in Eq.\,(\ref{eqn:cross-7}).

The second term corresponds to the cross section of Eq.\,(97) 
in Ref.\,\cite{ks-gluon-trans-2019}.
However, here we take the specific polarization $S^{xx}_{B,TT}$
instead of the combination $E_x - E_y$ with $S^{xx}_{B,TT} = -1$
of Ref.\,\cite{ks-gluon-trans-2019}. 
Hence, Eq.\,(97) of Ref.\,\cite{ks-gluon-trans-2019} needs 
to be multiplied by $-(1/2) S^{xx}_{B,TT}$, as shown 
in Eq.\,(\ref{eqn:cross-5}).

\begin{widetext}
The actual expression of the cross section is given for the first term as
\cite{ks-gluon-trans-2019}
\begin{align}
& \! \! \! \! \! \! \! \! 
\frac{ d \sigma_{0} (S_{B,LL}) }{d\tau \, dq_T^2 \, d\phi \, dy}
=  \frac{\alpha^2 \, \alpha_s \, C_F}{4 \, \pi \, \tau \, s^2} 
\int_{\text{min}(x_a)}^1 dx_a \frac{1}{(x_a -x_1) \, x_a^2 \, x_b^2}
\nonumber \\
& 
\! \! \! \! \! \! \! \! \! 
\times \sum_{q}  e_q^2 \, \bigg [ \frac{4}{9} 
\left\{ f_{1,q/A} (x_a) \, f\,'_{\! \! 1,\bar q/B} (x_b) 
      + f_{1,\bar q/A} (x_a) \, f\,'_{\! \! 1,q/B} (x_b) \right\}
\frac{2 \tau \left\{\tau -(-2 x_a x_b +x_1 x_b +x_2 x_a) \right\} 
                    +x_b^2 (x_a-x_1)^2+x_a^2 (x_b-x_2)^2 }{(x_a-x_1)(x_b-x_2)}
\nonumber \\
& \ \ \ \ \ \ \ \ \ \ \ \ 
 + \frac{1}{6} \left\{ f_{1,q/A} (x_a) + f_{1, \bar q/A} (x_a) \right\} 
                   f\,' _{\! \! 1,g/B} (x_b)
     \frac{ 2\tau (\tau-x_1 x_b) +x_b^2 \left\{ (x_a-x_1)^2 + x_a^2 \right\}}
          {x_b (x_a-x_1)}
\nonumber \\
& \ \ \ \ \ \ \ \ \ \ \ \ 
 + \frac{1}{6} \, f_{1,g/A} (x_a) \left\{ f\, ' _{\! \! 1,q/B} (x_b) 
                                        + f\, ' _{\! \! 1,\bar q/B} (x_b) \right\} 
    \frac{ 2\tau (\tau-x_2 x_a) + x_a^2 \left\{ (x_b-x_2)^2 + x_b^2 \right\}}
         {x_a (x_b-x_2)}
 \bigg ],
\label{eqn:cross-7}
\end{align}
where the parton distribution $f\,'_{\! \! 1,X/B} (x_b)$ 
is defined as
$ f\,'_{\! \! 1,X/B} (x_b) =  f_{1,X/B} (x_b ) 
                            +S_{B,LL} \, f_{1LL,X/B} (x_b) $.
In the numerical analysis of Ref.\,\cite{ks-gluon-trans-2019},
the tensor-polarized PDFs $f_{1LL,X/B} (x_b)$ are neglected
because they are considered to be very small in comparison
with the unpolarized PDFs.
The variables $x_1$ and $x_2$ are given by
the transverse mass $M_T = \sqrt{Q^2 + \vec q_T^{\,\, 2}}$,
the rapidity $y$, and the c.m. energy squared $s$ as
$x_1  = M_T \, e^{\,y} / \sqrt{s}$ and $x_2  = M_T \, e^{-y} / \sqrt{s}$.
The momentum fraction $x_b$ and $\text{min}(x_a)$ are expressed by
these variables as $x_b= (x_a x_2 - \tau)/(x_a - \tau)$ and
$\text{min}(x_a) = (x_1-\tau)/(1-x_2)$.
The second cross-section term is given by \cite{ks-gluon-trans-2019}
\begin{align}
\frac{ d \sigma_{\Delta_T g} (S_{B,TT}^{xx}) }{d\tau \, dq_T^2 \, d\phi \, dy} 
= \frac{\alpha^2 \, \alpha_s \, C_F \, q_T^2}{12 \pi s^3}
 \, S_{B,TT}^{xx} \, \cos (2\phi) 
 \int_{\text{min}(x_a)}^1 & dx_a
 \frac{1} { (x_a x_b)^2 \, (x_a -x_1) (\tau -x_a x_2 )^2} 
\nonumber \\
&  
\times 
 \sum_{q}  e_q^2 \, x_a
 \left[ \, f_{1,q/A} (x_a) + f_{1,\bar q/A} (x_a) \, \right ]
  x_b \Delta_T g_B (x_b) .
\label{eqn:cross-5}
\end{align}

Now, let us consider the two rotations for the deuteron polarization
in the laboratory frame with the deuteron at rest.
One is to take the rotation angle $\alpha_y = \alpha$ around 
the $y$ axis, and the other is to take the rotation 
angle $\alpha_x = \alpha$ around the $x$ axis
as shown in Fig.\,\ref{fig:lab-spin}.
Then, if we take the difference of the cross sections at the two angles, 
the first cross-section term 
$d \sigma_{0} (S_{LL}) / (d\tau \, dq_T^2 \, d\phi \, dy)$ 
drops and we obtain
\begin{align}
\frac{ d \sigma_{pd \to \mu^+ \mu^- X} (\alpha_y = \alpha) }{d\tau \, dq_T^2 \, d\phi \, dy}
& -\frac{ d \sigma_{pd \to \mu^+ \mu^- X} (\alpha_x = \alpha) }{d\tau \, dq_T^2 \, d\phi \, dy}
= \frac{\alpha^2 \, \alpha_s \, C_F \, q_T^2}{12 \pi s^3}
 \, \sin^2 \alpha \, \cos (2\phi) 
\nonumber \\
&  \! \! \! \! \! \! \! \! \! \! \! \! \! \! 
\times \int_{\text{min}(x_a)}^1 dx_a 
 \frac{1} { (x_a x_b)^2 \, (x_a -x_1) (\tau -x_a x_2 )^2} 
 \sum_{q}  e_q^2 \, x_a
 \left[ \, f_{1, q/A} (x_a) + f_{1, \bar q/A} (x_a) \, \right ]
  x_b \Delta_T g_B (x_b) .
\label{eqn:cross-pd-rotate-minus}
\end{align}
On the other hand, their cross-section summation is given by
\begin{align}
& \frac{ d \sigma_{pd \to \mu^+ \mu^- X} (\alpha_y = \alpha) }{d\tau \, dq_T^2 \, d\phi \, dy}
 +\frac{ d \sigma_{pd \to \mu^+ \mu^- X} (\alpha_x = \alpha) }{d\tau \, dq_T^2 \, d\phi \, dy}
=  \frac{\alpha^2 \, \alpha_s \, C_F}{ 2 \, \pi \, \tau \, s^2} 
\int_{\text{min}(x_a)}^1 dx_a \frac{1}{(x_a -x_1) \, x_a^2 \, x_b^2}
\nonumber \\
& \! \! \! \! \!
\times \sum_{q}  e_q^2 \, \bigg [ \frac{4}{9} 
\left\{ f_{1,q/A} (x_a) \, f\,' _{\! \! 1,\bar q/B} (x_b) 
       +f_{1,\bar q/A} (x_a) \, f\,'_{\! \! 1,q/B} (x_b) \right\}
\frac{2 \tau \left\{\tau -(-2 x_a x_b +x_1 x_b +x_2 x_a) \right\} 
                    +x_b^2 (x_a-x_1)^2+x_a^2 (x_b-x_2)^2 }{(x_a-x_1)(x_b-x_2)}
\nonumber \\
& \ \ \ \ \ \ \ \ \ \ \ \ 
 + \frac{1}{6} \left\{ f_{1,q/A} (x_a) + f_{1, \bar q/A} (x_a) \right\} 
            f\, ' _{\! \! 1,g/B} (x_b)
     \frac{ 2\tau (\tau-x_1 x_b) +x_b^2 \left\{ (x_a-x_1)^2 + x_a^2 \right\}}
          {x_b (x_a-x_1)}
\nonumber \\
& \ \ \ \ \ \ \ \ \ \ \ \ 
 + \frac{1}{6} \, f_{1,g/A} (x_a) \left\{ f\, ' _{\! \! 1, q/B} (x_b) 
                               + f\, ' _{\! \! 1, \bar q/B} (x_b) \right\} 
    \frac{ 2\tau (\tau-x_2 x_a) + x_a^2 \left\{ (x_b-x_2)^2 + x_b^2 \right\}}
         {x_a (x_b-x_2)}
 \bigg ],
\label{eqn:cross-pd-rotate-plus}
\end{align}
with the PDFs 
$ f\,'_{\! \! 1,X/B} (x_b ) = f_{1,X/B} (x_b ) + \frac{1}{8} 
\left\{ 1 \! + \! 3\cos (2\alpha) \right\} \, f_{1LL,X/B}  (x_b)$.
\end{widetext}

Therefore, rotating the deuteron longitudinal polarization
$\vec s_{+1}$ around the transverse coordinates $x$ and $y$
as shown in Fig.\,\ref{fig:lab-spin},
we can measure the gluon transversity as the difference
between the two cross sections $d\sigma (\alpha_x=\alpha)$
and $d\sigma (\alpha_y=\alpha)$, whereas their summation
is given by the unpolarized and tensor-polarized PDFs
of the deuteron. 
If $\alpha=\pi/2$ is taken, the gluon transversity distribution
can be measured in the proton-deuteron Drell-Yan process
with the deuteron polarizations along the two-transverse 
directions. 
In the same way, rotations of the polarization$\vec E_-$ 
or $s_{-1}$ can be used for investigating
the gluon transversity distribution.

\section{Summary}\label{summary}

For finding the gluon transversity distribution of the deuteron in
the proton-deuteron Drell-Yan process, the linear polarizations
of the deuteron are needed theoretically as investigated in
Ref.\,\cite{ks-gluon-trans-2019}.
Since the linear polarizations are rarely used in handing the deuteron
experimentally, we showed in this work that the Drell-Yan cross sections 
are expressed by the usual deuteron spin polarizations 
by rotating the spin vector around the two transverse axes. 
Then, we indicated that the difference of the two cross sections 
can be used for finding the gluon transversity distribution in the deuteron.
With the transversely-polarized deuteron along two transverse directions, 
such a gluon transversity measurement is possible 
at hadron-accelerator facilities.

\begin{acknowledgements}

The authors thank W.~Cosyn, D.~Keller, and Y.~Miyachi for discussions
on deuteron polarizations. 
This work was partially supported by Japan Society for the Promotion 
of Science (JSPS) Grants-in-Aid for Scientific Research (KAKENHI) 
Grant Number 19K03830.
\end{acknowledgements}




\begin{thebibliography}{99}
\bibitem{nucleon-spin}
For review, see 
   S. E. Kuhn, J.-P. Chen, and E. Leader, 
     Prog. Part. Nucl. Phys. {\bf 63}, 1 (2009);
   A. Deur, S. J. Brodsky, and G. F. de Teramond, 
     Rep. Prog. Phys. {\bf 82}, 076201 (2019),
   and references therein.
\bibitem{decomposition}
    E. Leader and C. Lorce, Phys. Rep. {\bf 541}, 163 (2014);
    M. Wakamatsu, Int. J. Mod. Phys. A {\bf 29}, 1430012 (2014).
\bibitem{lattice-pdfs}
    X. Ji, Phys. Rev. Lett. {\bf 110}, 262002 (2013).
    For recent progress, 
    see $e.g.$ T. Ishikawa {\it et al.}, Phys. Rev. D {\bf 96}, 094019 (2017);
    Huey-Wen Lin {\it et al.}, Prog. Part. Nucl. Phys. {\bf 100}, 107 (2018);
    Yu-Sheng Liu {\it et al.}, Phys. Rev. D {\bf 101}, 034020 (2020). 
\bibitem{gpds-gdas}
  M.~Diehl, Phys. Rep.  {\bf 388}, 41 (2003);
  S.~Wallon, Doctoral school lecture notes on courses ED-107 and ED-517,
     Universit\'e Paris Sud (2014), unpublished.
\bibitem{gpds}  
  K.~Goeke, M.~V.~Polyakov, and M.~Vanderhaeghen,
  Prog. Part. Nucl. Phys. {\bf 47}, 401 (2001);
  X.~Ji,  Annu. Rev. Nucl. Part. Sci. {\bf 54}, 413 (2004);
  A.~V.~Belitsky and A.~V.~Radyushkin, Phys. Rep.  {\bf 418}, 1 (2005);
  S.~Boffi and B.~Pasquini,  Riv. Nuovo Cimento  {\bf 30}, 387 (2007);
    M. Diehl and P. Kroll, Eur. Phys. J. C {\bf 73}, 2397 (2013);
    D. Mueller, Few Body Syst. {\bf 55},  317 (2014); 
	K. Kumericki, S. Liuti, and H. Moutarde,
                Eur. Phys. J. A {\bf 52}, 157 (2016);
    H. Moutarde, P. Sznajder, and J. Wagner, 
                Eur. Phys. J. C {\bf 78}, 890 (2018).
\bibitem{gdas}
    S. Kumano, Qin-Tao Song, and O. V. Teryaev,  
           Phys. Rev. D {\bf 97}, 014020 (2018).
\bibitem{tmds}
  U.~D'Alesio and F.~Murgia, Prog. Part. Nucl. Phys. {\bf 61}, 394 (2008);
  V.~Barone, F.~Bradamante, and A.~Martin,
      Prog. Part. Nucl. Phys. {\bf 65}, 267 (2010);
  C.~A.~Aidala, S.~D.~Bass, D.~Hasch, and G.~K.~Mallot,
      Rev. Mod. Phys. {\bf 85}, 655 (2013);
  M.~G.~Perdekamp and F.~Yuan,
      Annu. Rev. Nucl. Part. Sci. {\bf 65}, 429 (2015).
\bibitem{br-book} 
   V. Barone and R. G. Ratcliffe, 
       {\it Transverse Spin Physics} (World Scientific, Singapore, 2003).     
\bibitem{transversity-pdfs}
   Z.-B. Kang, A. Prokudin, P. Sun, and F. Yuan, 
         Phys. Rev. D {\bf 93}, 014009 (2016);
    M. Radici and A. Bacchetta, Phys. Rev. Lett. {\bf 120}, 192001 (2018);
    J. Cammarota {\it et al.}, arXiv:2002.08384.
\bibitem{gluon-trans-th}
    R. L. Jaffe and A. Manohar, Phys. Lett. {\bf B223}, 218 (1989);
    J. P. Ma, C. Wang, and G. P. Zhang, arXiv:1306.6693 (unpublished).
\bibitem{transversity-model}
    M.~Nzar and P.~Hoodbhoy,
         Phys. Rev. D {\bf 45}, 2264 (1992).
\bibitem{transversity-lattice}
   W. Detmold and P. E. Shanahan, Phys. Rev. D {\bf 94}, 014507 (2016);
                                            {\bf 95}, 079902 (2017).
\bibitem{transversity-q2}
   F. Baldracchini, N. S. Craigie, V. Roberto, and M. Socolovsky,
       Fortsch. Phys. {\bf 30}, 505 (1981);
   X. Artru and M. Mekhfi, Z. Phys. C {\bf 45}, 669 (1990);
   S. Kumano and M. Miyama, Phys. Rev. D {\bf 56}, R2504 (1997);
   A. Hayashigaki, Y. Kanazawa, and Y. Koike, Phys. Rev. D {\bf 56}, 7350 (1997);
   W. Vogelsang, Phys. Rev. D {\bf 57}, 1886 (1998);
   M. Hirai, S. Kumano, and M. Miyama, Comput. Phys. Commun. {\bf 111}, 150 (1998).
\bibitem{transversity-q2-gluon}
   W. Vogelsang, Acta Phys. Pol. B {\bf 29}, 1189 (1998).
\bibitem{b1} P. Hoodbhoy, R. L. Jaffe, and A. Manohar,
                   Nucl. Phys. {\bf B312}, 571 (1989);
   R. L. Jaffe and A. Manohar, Nucl. Phys. {\bf B321}, 343 (1989);
   F. E. Close and S. Kumano, Phys. Rev. D  {\bf 42}, 2377 (1990);
   S. Hino and S. Kumano, Phys. Rev. D {\bf 59}, 094026 (1999);
                                       {\bf 60}, 054018 (1999);       
   T.-Y. Kimura and S. Kumano, Phys. Rev. D {\bf 78}, 117505 (2008);
   S. Kumano, Phys. Rev. D {\bf 82}, 017501 (2010);
                  J. Phys.: Conf. Series {\bf 543}, 012001 (2014);
   G. A. Miller,  Phys. Rev. C {\bf 89}, 045203 (2014).
\bibitem{b1-convolution} 
   W. Cosyn, Yu-Bing Dong, S. Kumano, and M. Sargsian,
                 Phys. Rev. D {\bf 95}, 074036 (2017). 
\bibitem{b1-hermes} 
   A. Airapetian {\it et al.} (HERMES Collaboration), 
                 Phys. Rev. Lett. {\bf 95}, 242001 (2005).
\bibitem{Jlab-b1} Proposal to Jefferson Lab PAC-38, 
                         J.-P. Chen {\it et al.} (2011);
     K. Slifer, E. Long, and J. Maxwell,
     talks at the Workshop on Exploring QCD with light nuclei at EIC,
     Stony Brook, New York, USA,
     https://indico.bnl.gov/event/6799/ .
\bibitem{Fermilab-dy} 
    Fermilab E1039 experiment, Letter of Intent Report No. P1039 (2013), 
      https://www.fnal.gov/directorate\\ 
      /program\_planning/June2013PACPublic/P-1039\_LOI \\ \_polarized\_DY.pdf.
    For the ongoing Fermilab E-906 \\/SeaQuest experiment, see
           http://www.phy.anl.gov/mep \\ /drell-yan/.     
\bibitem{ks-2016}
    S. Kumano and Qin-Tao Song, Phys. Rev. D {\bf 94}, 054022 (2016). 
\bibitem{jlab-gluon-trans}
    A Letter of Intent to Jefferson Lab PAC 44, 
         LOI12-16-006, M. Jones {\it et al.} (2016);
    R.~L.~Jaffe and A.~Manohar, Phys. Lett. B {\bf 223}, 218 (1989);
    E.~Sather and C.~Schmidt, Phys. Rev. D {\bf 42}, 1424 (1990);
    J. P. Ma, C. Wang, and G. P. Zhang, arXiv:1306.6693 (unpublished).
\bibitem{j-parc} 
   J-PARC hadron project: https://j-parc.jp/Hadron/en/.
\bibitem{gsi-fair} 
   GSI-FAIR project: https://fair-center.eu/.
\bibitem{NICA-SPD}
    For the Spin Physics Detector (SPD) project at NICA, see http://spd.jinr.ru/.
\bibitem{RHIC-fixed}
    D. Cebra, personal communications on the RHIC fixed-target project (2020).
\bibitem{ks-gluon-trans-2019}
    S. Kumano and Qin-Tao Song, Phys. Rev. D {\bf 101}, 054011 (2020).
\bibitem{photon-polarization}
   M. Born and E. Wolf, {\it Principles of optics} 
         (Cambridge University Press, Cambridge, 1999);
   J. D. Jackson, {\it Classical Electrodynamics}
         (John Wiley \& Sons, Inc., New York, 1975).
\bibitem{leader-book}
    E. Leader, {\it Spin in Particle Physics} 
     (Cambridge University Press, Cambridge, 2001).
\bibitem{bacchetta-2000-PRD}
    A. Bacchetta and P. J. Mulders, Phys. Rev. D {\bf 62}, 114004 (2000).
\bibitem{vonDaal-2016}
    T. van Daal, arXiv:1612.06585; arXiv:1812.07336, 
           Ph. D. thesis, University of Groningen (2018).
\bibitem{Boer-2016}
    D. Boer {\it et al.}, J. High Energy Phys. {\bf 10}, 013 (2016).
\end{thebibliography}
\end{document}